\documentclass[12pt]{article}
\usepackage{graphics}
\usepackage{amssymb}

\renewcommand{\theequation}{\arabic{section}.\arabic{equation}}

\newcommand{\e}{\mathrm{e}}
\newcommand{\C}{\mathbb{C}}
\newcommand{\R}{\mathbb{R}}
\newtheorem{claim}{Claim}[section]
\newtheorem{theorem}[claim]{Theorem}
\newtheorem{lemma}[claim]{Lemma}
\newtheorem{remark}[claim]{Remark}
\newtheorem{remarks}[claim]{Remarks}
\newenvironment{proof}[1][Proof]{\textsl{#1.} }{\ \rule{0.5em}{0.5em}}

\begin{document}

\title{Curvature-induced bound states for a $\delta$ interaction
supported by a curve in $\R^3$}
\author{P.~Exner and S.~Kondej}
\date{}
\maketitle

\begin{quote}
{\small {\bf Abstract.} We study the Laplacian in $L^2(\R^3)$
perturbed on an infinite curve $\Gamma$ by a $\delta$ interaction
defined through boundary conditions which relate the corresponding
generalized boundary values. We show that if $\Gamma$ is smooth
and not a straight line but it is asymptotically straight in a
suitable sense, and if the interaction does not vary along the
curve, the perturbed operator has at least one isolated eigenvalue
below the threshold of the essential spectrum.}
\end{quote}


\section{Introduction}

Relations between the geometry and spectral properties are one of
the vintage topics of mathematical physics. In the last decade
they attracted attention also in the context of quantum mechanics.
A prominent example is the curvature-induced binding in infinite
tube like regions \cite{ES, GJ, DE, RB}. This effect appears to be
a robust one: it has been demonstrated recently that bends can
produce localized states not only if the transverse confinement is
hard, i.e. realized by a Dirichlet condition, but also when it is
weaker corresponding to a potential well or a $\delta$ interaction
\cite{EI}.

The result is appealing, not only because it concerns an
interesting mathematical problem, but also in view of applications
in mesoscopic physics where such operators are used as a natural
model for semiconductor ``quantum wires''. Since in the latter
electrons are trapped due to interfaces between two different
materials representing finite potential jumps, by tunneling effect
they can be found outside the wire, albeit not too far because the
exterior is (for the energies in question) the classically
forbidden region.

The main result of the paper \cite{EI} concerns nontriviality of
the discrete spectrum for a class of operators in $L^2(\R^2)$
which can be formally written as $-\Delta-
\alpha\delta(x\!-\!\Gamma)$ with $\alpha>0$, where $\Gamma$ is a
curve which is not a straight line but it is asymptotically
straight in a suitable sense. A question naturally arises whether
a similar result is valid for a curve in $\R^3$. Such an extension
is not trivial, because the argument in \cite{EI} relies on the
resolvent formula of \cite{BEKS} representing in a sense a
generalization of the Birman-Schwinger theory. The said formula is
valid for singular perturbations of the Laplacian which can be
treated by means of a quadratic-form sum, i.e. as long as the
codimension of the manifold supporting the perturbation is one.

Thus if we want to address the stated question, we are forced to
look for other tools. One possibility is to employ the resolvent
formula for a curve in $\R^3$ derived in \cite{Ku}. However, since
it uses rather strong regularity hypotheses about the curve we
take another route and begin instead with an abstract formula for
strongly singular perturbations due to A.~Posilicano \cite{AP}.
When it is specified to our particular case, it contains again an
embedding operator into a space of functions supported on the
curve $\Gamma$, however, this time it is not the ``naive'' $L^2$
but rather a suitable element from the scale of Sobolev spaces. Of
course, one can regard it as a generalization of Krein's formula;
recall that such a way of expressing the resolvent can be used not
only to describe $\delta$ interaction perturbations but also more
general dynamics supported by zero measure sets \cite{Ka, KK, Ko}.

Another aspect of the absence of a description in terms of the
quadratic-form sum concerns the very definition of the operator we
want to study. We have to employ boundary conditions which relate
the corresponding generalized boundary values in the normal plane
to the curve modeled after the usual two-dimensional $\delta$
interaction \cite{AGHH}, which requires us to impose stronger
regularity conditions on $\Gamma$. Furthermore, a modification of
the Birman-Schwinger technique used in \cite{EI} demands stronger
restrictions on the regularity of the curve. On the other hand,
apart of these technical hypotheses our main result -- stated in
Theorem~\ref{ispo10} below -- is analogous to that of \cite{EI},
namely that for any curve which is asymptotically straight but not
a straight line the corresponding operator has at least one
isolated eigenvalue. This conclusion is by no means obvious having
in mind how different are the point interactions in one and two
dimensions.


\section{The resolvent formula}
\setcounter{equation}{0}

As a preliminary let us show how self-adjoint extensions of
symmetric operators can be characterized in terms of a Krein-type
formula derived in \cite{AP}; we refer to this paper for the proof
and a more detailed discussion. With a later purpose on mind we do
not strive for generality and restrict ourselves to the case of
the Hilbert space $\mathcal{H}:=L^{2}(\R^3)\equiv L^{2}$ and the
Laplace operator, $-\Delta :D(\Delta )\to L^{2}$, which is well
known to be self-adjoint on the domain $D(\Delta )$ which
coincides with the usual Sobolev space $H^{2}(\R^3)\equiv H^{2}$.

For any $z$ belonging to the resolvent set $\varrho (-\Delta )
=\C\setminus \lbrack 0,\infty )$ we define the resolvent as the
bounded operator $R^{z}:=(-\Delta -z)^{-1}:L^{2}\to H^{2}$.
Consider a bounded operator
 $$ 
 \tau :H^{2}\to \mathcal{X}
 $$ 
into a complex Banach space $\mathcal{X}$ and its adjoint in the
dual space $\mathcal{X}^{\prime }$. Recall that for a closed
linear operator $A:\mathcal{X}\to \mathcal{Y}$ the adjoint is
defined by $(A^{\ast }l)(x)=l(Ax)$ for all $x\in D(A)$ and $l\in
D(A^{\ast })\subseteq \mathcal{Y}^{\prime }$. Then we can
introduce the operators
 $$ 
 R_{\tau}^{z}=\tau R^{z}:L^{2}\to \mathcal{X}\,, \quad
 \breve{R}_{\tau}^{z}=(R_{\tau }^{\bar{z}})^{\ast }:
 \mathcal{X}^{\prime }\to L^{2}\,,
 $$ 
which are obviously bounded too. Let $Z$ be an open subset of
$\varrho (-\Delta )$ symmetric w.r.t. the real axis, i.e. such
that $z\in Z$ implies $\bar{z}\in Z$. Suppose that for any $z\in
Z$ there exists a closed operator $Q^{z}:D\subseteq
\mathcal{X}^{\prime }\to \mathcal{X}$ satisfying the following
conditions,
 \begin{eqnarray}
 && Q^{z}-Q^{w}=(z\!-\!w)R_{\tau }^{w}\breve{R}_{\tau }^{z}\,,
 \label{condi1} \\ &&
 \forall l_{1},l_{2}\in D\,,\quad  \ l_{1}(Q^{\bar{z}}
 l_{2})=\overline {l_{2}(Q^{z}l_{1})}\,.  \label{condi2}
 \end{eqnarray}
It will be used to construct a family of self-adjoint operators
which coincide with $-\Delta $ when restricted to $\ker \tau $.
They can be parametrized by symmetric operators $\Theta :D(\Theta
)\subseteq \mathcal{X}^{\prime }\to \mathcal{X}$. To this end, we
define
 \begin{eqnarray*}
 && Q_{\Theta }^{z}=\Theta +Q^{z}:D(\Theta )\cap D \subseteq
 \mathcal{X} ^{\prime }\to \mathcal{X}\,, \\ &&
 Z_{\Theta }:= \{\,z\in \rho (-\Delta ):\,(Q_{\Theta }^{z})^{-1},\,
 (Q_{\Theta}^{\bar{z} })^{-1} \; \mathrm{exist\ and\ are\
 bounded}\,\}\,.
 \end{eqnarray*}
With this notation we can state the result we want to borrow from
\cite{AP}.

\begin{theorem} \label{Posil}
Assume that the conditions
 \begin{equation}
 Z_{\Theta } \neq \emptyset  \label{hipot1}
 \end{equation}
and
 \begin{equation}
 \mathrm{Ran}\, \tau ^{\ast }\cap L^{2} =\{0\}  \label{hipot2}
 \end{equation}
are satisfied. Then the bounded operator
 $$ 
 R_{\tau ,\Theta }^{z}:= R^{z}-\breve{R}_{\tau }^{z}
 (Q_{\Theta}^{z})^{-1}R_{\tau }^{z}\,,\quad z\in Z_{\Theta }\,,
 $$ 
is the resolvent of the self-adjoint operator $-\Delta
_{\tau,\Theta }$ defined by
 \begin{eqnarray*}
 && D(\Delta _{\tau ,\Theta })=\{\,f\in L^{2}:f=f_{z}
 \!-\!\breve{R}_{\tau}^{z}(Q_{\Theta }^{z})^{-1}\tau f_{z},\
 f_{z}\in D(\Delta )\,\}\,, \\ &&
 (-\Delta _{\tau ,\Theta }-z)f :=(-\Delta -z)f_{z}\,,
 \end{eqnarray*}
which coincides with $-\Delta $ on the $\ker \tau $.
\end{theorem}


\section{Singular perturbation on a curve in $\R^3$}
 \setcounter{equation}{0}

Henceforth, we will be interested in a specific class of
perturbations of the Laplacian on $\mathcal{H}=L^{2}(\R^{3})$. The
free resolvent
 $$ 
 R^{z}=(-\Delta -z)^{-1}:\: L^{2}(\R^{3})\to H^{2}(\R^{3})\,,
 \quad z\in \varrho (-\Delta )\,,
 $$ 
is an integral operator with the kernel
 $$ 
 G^{z}(x\!-\!y)=
 \frac{\e^{i\sqrt{z}\left| x-y\right| }}
 {4\pi \left| x\!-\!y\right| }\,.
 $$ 
Let $\Gamma \subset \R^{3}$ be a curve defined as a graph of a
continuous function which is assumed to be piecewise $C^{1}$.
Recall that $\Gamma$ admits a natural parametrization by the arc
length which is unique up to a choice if the reference point; we
denote the parameter as $s$ and use the symbol $\gamma (s):\:\R\to
\R^{3}$ for the corresponding function. Then we have
 \begin{equation}
 \left| \gamma (s)\!-\!\gamma (s^{\prime })\right| \leq
 \left|s\!-\!s^{\prime }\right|\, .  \label{lipshi}
\end{equation}
To specify further the family of curves which we will consider, we
introduce for any $\tilde{\omega}\in (0,1)$ and
$\tilde{\varepsilon}>0$ the set
 \begin{eqnarray*}
 S_{\tilde{\omega},\tilde{\varepsilon}} := \bigg\{\,
 (s,s^{\prime}):\: \tilde{\omega}< \frac{s}{s^{\prime }}
 <\tilde{\omega}^{-1} &\mathrm{if}&
 \left| s\!+\! s^{\prime}\right| >\xi
 (\tilde{\omega})\tilde{\varepsilon}\,, \\ \mathrm{and}\;
 \left|s\!-\!s^{\prime}\right| <\tilde{\varepsilon}
 &\mathrm{if}& \left| s\!+\!s^{\prime}\right| <\xi
 (\tilde{\omega}) \tilde{\varepsilon}\, \bigg\}\,,
 \end{eqnarray*}
where $\xi(\tilde{\omega}) := \frac{1+\tilde{\omega}}
{1-\tilde{\omega}}\,$. We adopt the following assumptions:

 \begin{description}
 \item{(a1)} there exists a $c\in (0,1)$ such that $\left|
 \gamma (s)\!-\!\gamma (s^{\prime})\right| \geq
 c\left| s\!-\!s^{\prime }\right| $,
 \vspace{-0.8ex}
 \item{(a2)} there are $\omega \in (0,1)$, $\mu \geq 0$
 and positive $\varepsilon,d$ such that the inequality
 $$ 
 1-\frac{\left| \gamma (s)\!-\!\gamma (s^{\prime })\right|}
 {\left| s\!-\!s^{\prime }\right| }\leq d\:\frac{\left|
 s\!-\!s^{\prime}\right| }{(\left| s\!-\!s^{\prime }\right| +1)
 (1+(s^{2}\!+\!s^{\prime 2})^{\mu })^{1/2}}
 $$ 
 holds for all $(s,s^{\prime })\in S_{\omega ,\varepsilon }$.
 \end{description}
The first condition means, in particular, that $\Gamma$ has no
cusps and self-intersections. The second assumption is basically a
requirement of asymptotic straightness (see Remark~\ref{a2
meaning}), but in contrast to \cite{EI} it restricts also the
behaviour of $|\gamma (s)\!-\!\gamma (s^{\prime })|$ at small
distances; it is straightforward to check that the bound cannot be
satisfied unless $\Gamma$ is $C^1$-smooth.

To make use of Theorem~\ref{Posil} we take $\mathcal{X}=L^{2}(\R)$
and denote the corresponding scalar product by $(\cdot
,\cdot)_{l}\:$ (see also Remark~\ref{trueX} below). The operator
$\tau :H^{2}(\R^{3})\to L^{2}(\R)$ which we will employ in our
construction is a trace map defined in the following way:
 $$ 
 \tau \phi (s):=\phi (\gamma(s))\,;
 $$ 
it is a standard matter to check that the definition makes sense
and the operator $\tau$ is bounded \cite{BN}. The adjoint operator
$\tau ^{\ast }:L^{2}(\R)\to H^{-2}(\R^{3})$ is determined by the
relation
 $$ 
 \left\langle \tau ^{\ast }h,\omega \right\rangle =(h,\tau
 \omega )_{l}\,, \quad h\in L^{2}(\R)\,,\quad \omega \in
 H^{-2}(\R^{3})\,,
 $$ 
where $\left\langle \cdot ,\cdot \right\rangle $ stands for the
duality between $H^{-2}(\R^{3})$ and $H^{2}(\R^{3})$, in other
words, we can write
 $$ 
 \tau ^{\ast }h=h\delta _{\Gamma }\,,
 $$ 
where $\delta _{\Gamma }$ is the Dirac measure supported by
$\Gamma $. Since $\delta _{\Gamma }\notin L^{2}(\R^{3})$ we get
 $$ 
 \mathrm{Ran\,}\tau ^{\ast }\cap L^{2}(\R^{3})=\{0\}\,,
 $$ 
so condition (\ref{hipot2}) is satisfied.

\begin{remark} \label{trueX}
{\rm Notice that the map $\tau $ as introduced above is not
surjective. Indeed, since $\gamma (s)$ is a Lipschitz function we
have $\mathrm{Ran\,}\tau =H^{1}(\R)$ -- cf.~\cite{BN}. However, we
lose nothing by keeping $\mathcal{X}=L^{2}(\R)$ in the further
discussion. }
\end{remark}

\noindent The problem at hand is to define an operator
$Q^{z}:D\subseteq L^{2}(\R)\to L^{2}(\R)$ satisfying the
conditions (\ref{condi1}) and (\ref{condi2}). To this end some
preliminaries are needed. Since our considerations concern
spectral properties at the negative halfline, it suffices for
further discussion to restrict ourselves to $z=-\kappa ^{2}$ with
$\kappa>0$. In such a case it is convenient to modify slightly the
used notation by introducing
 $$ 
 \mathbf{Q}^{\kappa }:= Q^{-\kappa ^{2}}\,,\quad
 \mathbf{R}_{\tau}^{\kappa }:= R_{\tau }^{-\kappa ^{2}}\,,\quad
 \mathbf{\breve{R}}_{\tau }^{\kappa }:= \breve{R}_{\tau }^{-\kappa ^{2}}\,.
 $$ 
and similarly
 $$ 
 \mathbf{G}^{\kappa }(s\!-\!s^{\prime} ):=
 \frac{\e^{-\kappa \left| s-s^{\prime }\right| }}{4\pi \left|
 s\!-\!s^{\prime }\right| }\,,\quad \mathbf{G}^{\kappa }( \gamma
 (s)\!-\!\gamma (s^{\prime }) )=\frac{ \e^{-\kappa \left| \gamma
 (s)-\gamma (s^{\prime })\right| }}{4\pi \left| \gamma (s)\!-\!\gamma
 (s^{\prime })\right| }\,.
 $$ 
The difference of these two kernels,
 $$ 
 B_{\kappa }(s,s^{\prime })= \mathbf{G}^{\kappa }(\gamma
 (s)\!-\!\gamma (s^{\prime }) )- \mathbf{G}^{\kappa }(
 s\!-\!s^{\prime })\,,
 $$ 
defines the integral operator $B_{\kappa }:\:D(B_{\kappa })\to
L^{2}(\R)$ with the domain $D(B_{\kappa })=\{f\in
L^{2}(\R):\:B_{\kappa }f\in L^{2}(\R)\}$. A key observation is
that this operator has a definite sign: in view of (\ref{lipshi})
and of the fact that the function $\xi\mapsto \frac{\e^{-\kappa
\xi }}{\xi }$ decreases monotonically for $\kappa,\,\xi $
positive, we have
 \begin{equation}
 B_{\kappa }(s,s^{\prime })\geq 0\,. \label{nieroB}
 \end{equation}
The operator $B_{\kappa }$ is related obviously with the deviation
of $\Gamma $ from a straight line; below we shall demonstrate that properties
for a curve satisfying the assumptions (a1) and (a2) with any
$\mu \geq 0$ is bounded (see Remark~\ref{bondBk}).

Next we need to show how the free resolvent kernel behaves when
one of the three dimensions is integrated out. By a direct
computation one can show that for all $\kappa,\kappa^{\prime }>0$
and $f_{1},f_{2} \in L^{2}(\R)$ the following relation,
 \begin{eqnarray*}
 \lefteqn{ \int _{\R^{2}} f_{1}(s) \overline{f_{2}(s^{\prime
 })}\, [\mathbf{G}^{\kappa }(s\!-\!s^{\prime })
 -\mathbf{G}^{\kappa ^{\prime }}(s\!-\!s^{\prime })
 ]\,ds\,ds^{\prime }} \\ &&
 =\int _{\R^{2}} f_{1}(s) \overline{f_{2}(s^{\prime
 })}\, [\check{T}_{\kappa }(s\!-\!s^{\prime })
 -\check{T}_{\kappa ^{\prime }}(s\!-\!s^{\prime }
 )]\,ds\,ds^{\prime }\,,
 \end{eqnarray*}
is valid, where
 \begin{equation}
 \check{T}_{\kappa }(s\!-\!s^{\prime }) :=
 -\frac{1}{(2\pi)^{2} }\int _{\R} \ln \left( p^{2}\!+\!\kappa^{2}
 \right) ^{1/2} \e^{ip(s-s^{\prime })}\,dp\,. \label{indepe}
 \end{equation}
This result means, in particular, that
 \begin{equation}
 \int _{\R^{2}} f_{1}(s)\overline {f_{2}(s^{\prime
 })}\, \mathbf{G}^{\kappa }(s\!-\!s^{\prime })\,ds\,ds^{\prime}
 -\int _{\R^{2}} f_{1}(s)\overline{f_{2}(s^{\prime })}\,
 \check{T}_{\kappa }(s\!-\!s^{\prime })\,ds\,ds^{\prime }
 \label{k-inde}
 \end{equation}
is $\kappa$-independent. Let $T_{\kappa }:\:D(T_{\kappa })\to
L^{2}(\R)$ be the integral operator with the domain $D(T_{\kappa
})=\{f\in L^{2}(\R):\int_{\R} \check{T}_{\kappa }( s\!-\!s^{\prime
}) f (s^{\prime })\, ds^{\prime }\in L^{2}(\R)\}$ and the kernel
$T_{\kappa }(s\!-\!s^{\prime }):= \check {T}_{\kappa }(
s\!-\!s^{\prime } )+\frac{1}{2 \pi }(\ln 2+\psi (1))$ where
$-\psi (1)\approx 0.577$ is Euler's number. Then $T_{k}$ is
self-adjoint and we can define the operator
 $$ 
 \mathbf{Q}^{\kappa }f =(T_{\kappa }\!+\!B_{\kappa })f:\: D\equiv
 D(T_{\kappa })\to L^{2}(\R)\,,
 $$ 
which is also self-adjoint and has the needed properties:

\begin{lemma}
The operators $Q^{-\kappa ^{2}}\equiv \mathbf{Q}^{\kappa }$
satisfy the conditions (\ref{condi1}), (\ref{condi2}).
\end{lemma}
\begin{proof}
Let $f_{1},f_{2}\in D$, then a direct computation yields
 \begin{eqnarray*}
 \lefteqn{(\kappa ^{2}-\kappa ^{\prime 2})(f_{1},\mathbf{R}_{\tau
 }^{\kappa ^{\prime }}\mathbf{\breve{R}}_{\tau }^{\kappa
 }f_{2})_{l} } \\ &&
 =\int_{\R^{2}}f_{1}(s) \overline {f_{2}(s^{\prime })}\, [
 \mathbf{G}^{\kappa }(\gamma (s)\!-\! \gamma (s^{\prime }))
 -\mathbf{G}^{\kappa ^{\prime }}(\gamma (s)\!-\!\gamma
 (s^{\prime }))]\,ds\,ds^{\prime }\,.
 \end{eqnarray*}
On the other hand, by definition of $\mathbf{Q}^{\kappa }$ and the
$\kappa$-independence of the expression (\ref{k-inde}) we find
that $(f_{1},(\mathbf{Q}^{\kappa }-\mathbf{Q}^{\kappa ^{\prime
}})f_{2})_{l}$ is also given by the right-hand side of the last
formula, which proves (\ref{condi1}). Since $\mathbf{Q}^{\kappa }$
is self-adjoint, the condition (\ref{condi2}) is satisfied too.
\end{proof} \vspace{1em}

\noindent The operator $\Theta :L^{2}(\R)\to L^{2}(\R)$ appearing
in Theorem~\ref{Posil} will be identified here with the
multiplication by a real number, $\Theta f=-\alpha f$ with $\alpha
\in \R$ and the sign convention made with a later purpose on mind.
Then the operator
 $$ 
 \mathbf{Q}_{\Theta }^{\kappa }=\Theta + \mathbf{Q}^{\kappa}:\:
 D\to L^{2}(\R)
 $$ 
is closed and by Proposition~1 of Ref.~\cite{AP} we conclude that
(\ref{hipot1}) is satisfied.

For simplicity we identify in the following the symbols of the
operators $\tau,\, \Theta $ with $\gamma,\, \alpha$, respectively.
In this notation Theorem~\ref{Posil} says the following: if
$\kappa \in Z_{\alpha}$, i.e. if the operator $(\mathbf{Q}_{\alpha
}^{\kappa })^{-1}=(\mathbf{Q}^{\kappa }\!-\!\alpha )^{-1}:\:
L^{2}(\R)\to L^{2}(\R)$ exists and is bounded, then
 \begin{equation}
 \mathbf{R}_{\gamma ,\alpha }^{\kappa }=\mathbf{R}^{\kappa}
 -\mathbf{\breve{R} }_{\gamma }^{\kappa }(\mathbf{Q}^{\kappa}
 \!-\!\alpha )^{-1}\mathbf{R}_{\gamma }^{\kappa }  \label{resolv}
 \end{equation}
is the resolvent of a self-adjoint operator which we denote as
$-\Delta _{\gamma ,\alpha }$. It coincides with $-\Delta $ on
$\ker \tau =\{g \in H^{2}(\R^{3}) :g(x)=0,\: x \in \Gamma \} $ and
 \begin{eqnarray*}
 && D(-\Delta _{\gamma ,\alpha }) =\{f\in L^{2}:\:
 f=f_{\kappa }-\mathbf{\breve{R}}_{\gamma }^{\kappa }
 (\mathbf{Q}^{\kappa }\!-\!\alpha )^{-1}\tau
 f_{k},\; f_{\kappa }\in D(\Delta )\}\,, \\ &&
 (-\Delta _{\gamma ,\alpha }+\kappa ^{2})f
 =(-\Delta +\kappa ^{2})f_{\kappa}\,.
 \end{eqnarray*}


\section{The interaction in terms of boundary \\ conditions}
\label{bc}
 \setcounter{equation}{0}

To proceed further we have to impose slightly stronger regularity
requirement on the curve $\Gamma$ . Specifically, we assume that
it is given by a function $\gamma (s):\:\R\to \R^{3}$ which is
$C^1$ everywhere and piecewise $C^{2}$, and satisfies the
condition (a1). Then we can introduce, apart of a discrete set,
the Frenet's frame for $\Gamma $, i.e. the triple
$(t(s),b(s),n(s))$ of the tangent, binormal and normal vectors,
which are by assumption piecewise continuous functions of $s$.
Given $\xi ,\eta \in \R$ we denote $ r =(\xi ^{2}\!+\!\eta
^{2})^{1/2}$ and define the set the ``shifted'' curve
 $$ 
 \Gamma _{r}\equiv \Gamma ^{\xi \eta } _{r}:= \{\, \gamma _{r}
(s)\equiv \gamma ^{\xi \eta } _{r}(s):=\gamma (s)+\xi b(s)+\eta
n(s)\,\}\,.
 $$ 
It follows from the smoothness of $\gamma$ in combination with
(a1) that there exists an $r _{0}>0$ such that $\Gamma _{r}\cap
\Gamma =\emptyset $ holds for each $r < r _{0}$.

Since any function $f\in H_{loc}^{2}(\R^{3}\setminus \Gamma )$ is
continuous on $\R^{3}\setminus \Gamma$ its restriction to $\Gamma
_{r},\: r < r _{0}$ is well defined; we denote it as
${f\!\upharpoonright}_{\Gamma_{r}}(s)$. In fact, we can regard
${f\!\upharpoonright}_{\Gamma_{r}}(s)$ as a distribution from
$D^{\prime }(\R)$ with the parameter $r$.
We shall say that a function  $f\in H_{loc}^{2}(\R^{3}\setminus
\Gamma )\cap L^{2}(\R^{3})$\ belongs to $\Upsilon $ if the
following limits
 \begin{eqnarray*}
 \Xi (f)(s) &\!:=\!& -\lim_{r \to 0}\frac{1}{\ln r }
 {f\!\upharpoonright}_{\Gamma_{r}}(s)\,, \\
 \Omega (f)(s) &\!:=\!& \lim_{r \to 0}
 \left[{f\!\upharpoonright}_{\Gamma_{r}} (s)+\Xi (f)(s)\ln r\right] \,,
 \end{eqnarray*}
exist a.e. in $\R$, are independent of the direction ${1\over
r}(\xi, \eta)$, and define functions from $ L^{2}(\R)$. The limits
here are understood in the sense of the $D^{\prime }(\R)$
topology. With these prerequisites we are able now to characterize
the operator $-\Delta _{\gamma ,\alpha }$ discussed above in terms
of (generalized) boundary conditions, postponing the proof to the
appendix.

\begin{theorem}  \label{bocon4}
With the assumption stated above we have
 \begin{eqnarray}
 D(-\Delta _{\gamma ,\alpha }) &\!=\!&
 \Upsilon _{\alpha }:= \{\,g\in \Upsilon :\:2\pi \alpha \Xi
 (g)(s)=\Omega (g)(s) \,\}\,, \label{bcond3} \\
 -\Delta_{\gamma , \alpha }f &\!=\!& -\Delta f \quad\mathrm{for}
 \quad x \in \R^{3}\setminus \Gamma\,. \nonumber
 \end{eqnarray}
\end{theorem}


\section{Curvature-induced bound states}
 \setcounter{equation}{0}

Let us first find the spectrum of $-\Delta _{\gamma _{0},\alpha }$
where $\gamma _{0}$ is a linear function describing a straight
line. Since $B_{\kappa }=0$ holds in this case we have
$\mathbf{Q}^{\kappa }=T_{\kappa }$. Then the resolvent formula
(\ref{resolv}) yields
 $$ 
 \sigma (-\Delta _{\gamma _{0},\alpha })=\{\,-\kappa ^{2}:\: \alpha \in
 \sigma (T_{\kappa })=\sigma _\mathrm{ac}(T_{\kappa })\,\}\,.
 $$ 
Using the momentum representation of $T_{\kappa }$ we immediately
get
 $$ 
 \sigma _\mathrm{ac}(T_{\kappa }) =(-\infty ,s_{\kappa }]\,,
 $$ 
where $s_{\kappa }:=\frac{1}{2 \pi }(\psi (1)-\ln (\kappa /2))$.
Hence the spectrum of $ -\Delta _{\gamma _{0}, \alpha }$ is given
by
 $$ 
 \sigma (-\Delta _{\gamma _{0},\alpha })=\sigma _\mathrm{ac}
 (-\Delta_{\gamma _{0},\alpha })=[\zeta _{0},\infty )\,,
 $$ 
where $\zeta _{0}=-4 \e^{2(-2 \pi \alpha +\psi (1))}$ as we expect
with the spectrum of a two-dimen\-sional $\delta$ interaction
\cite{AGHH} and the natural separation of variables in mind.

To find the spectrum of $-\Delta _{\gamma ,\alpha }$ for a
non-straight curve we treat the respective operator
$\mathbf{Q}^{\kappa }$ as a perturbation of the one corresponding
to a straight line. First we have to localize the essential
spectrum. Following step by step the argument given in the proof
of Proposition~1 of Ref.~\cite{EI} we get

\begin{lemma} \label{espec5}
Let $\Gamma $ be a curve given by a function $\gamma (s)$
satisfying (a1) and (a2) with $\mu > 1/2$. Then $\sigma
_\mathrm{ess}(-\Delta _{\gamma ,\alpha })
 =[\zeta _{0},\infty )$.
\end{lemma}

Next we observe that a nontrivial bending pushes the upper bound
of the spectrum of $\mathbf{Q}^{\kappa }$ up.

\begin{lemma} \label{supsp6}
If $\Gamma $ is not a straight line we have
 \begin{equation}
 \sup \sigma (\mathbf{Q}^{\kappa })>s_{\kappa }\,. \label{supspe}
 \end{equation}
\end{lemma}
\begin{proof}
Let $\phi $ be a positive function from $C_{0}^{\infty}(\R^{3})$.
Given $\lambda>0 $ we set $\phi _{\lambda }(s):= \lambda
^{1/2}\phi (\lambda s)$. To show (\ref{supspe}) it suffices to
check the following inequality
 $$ 
 (\mathbf{Q}^{\kappa }\phi _{\lambda },\phi _{\lambda})_{l}
 -s_{\kappa }(\phi _{\lambda },\phi _{\lambda })_{l}>0\,,
 $$ 
which is easily seen to be equivalent to
 \begin{equation}
 -\frac{1}{2 \pi }\int_{\R} \ln \left( 1\!+\!\left(
 \frac{\lambda u}{\kappa }\right) ^{2}\right) ^{1/2}\left|
 \hat{\phi}(u)\right|^{2}du +\lambda \int_{\R^{2}}
 B_{\kappa }(s,s^{\prime })\phi (\lambda s) \phi (\lambda s^{\prime })\,
 ds\,ds^{\prime }>0\,, \label{sharp1}
 \end{equation}
where $\hat{\phi}$ stands for the Fourier transform of $\phi $.
The first term in the last expression can expanded as
 $$ 
 -\frac{1}{4 \pi }\left( \frac{\lambda }{\kappa }\right)^{2}
 \int_{\R}u^{2}\left| \hat{\phi}(u)\right|^{2}\,du
 +\mathcal{O}(\lambda ^{4})\,.
 $$ 
Since $\Gamma $ is not straight by assumption the inequality
(\ref{nieroB}) is sharp in an open subset of $\R^{2}$, so there is
$D>0$ such that $\lambda \int_{\R^{2}}B_{\kappa }(s,s^{\prime
})\phi (\lambda s)\phi (\lambda s^{\prime })\, ds\,ds^{\prime
}\geq D\lambda $ as $\lambda \to 0$. Consequently, for all
sufficiently small $\lambda $ the inequality (\ref{sharp1}) is
satisfied.
\end{proof} \vspace{1em}

On the other hand, the part of the spectrum in $(s_{\kappa},
\infty )$ added in this way is at most discrete provided the curve
has the asymptotic straightness properties expressed by the
assumption (a2) with $\mu$ large enough.

\begin{lemma} \label{HSnor7}
If $\mu >1/2$ then $B_{k}$ are Hilbert-Schmidt operators. Moreover, norms $%
\left\| B_{\kappa }\right\| _{HS}$ are uniformly bounded with
respect $\kappa \geq \kappa _{0}=\left| \zeta _{0}\right| ^{1/2}$.
\end{lemma}
\begin{proof}
Denote $\rho \equiv \rho (s,s^{\prime }):=\left| \gamma
(s)\!-\!\gamma (s^{\prime })\right|$ and $\sigma \equiv \sigma
(s,s^{\prime }):=\left| s\!-\!s^{\prime }\right|$. In this
notation the assumptions (a1), (a2) can be written as
 \begin{description}
 \item{(a1)} there is a $c\in (0,1)$ such that $\rho (s,s^{\prime })
 \geq c\sigma (s,s^{\prime })$,
 \vspace{-0.8ex}
 \item{(a2)} there are $\omega \in (0,1)$,
 $\mu \geq 0$ and $\varepsilon,d>0 $ s.t. for all
 $(s,s^{\prime })\in S_{\omega ,\varepsilon }$ we have
 $$ 
 1-\frac{\rho (s,s^{\prime })}{\sigma (s,s^{\prime})}\leq
 \frac{d\sigma (s,s^{\prime })}{(\sigma (s,s^{\prime})\!+\!1)
 (1+(s^{2}\!+\!s^{\prime 2})^{\mu })^{1/2}}\,.
 $$ 
 \end{description}
Next we notice that the perturbation kernel is monotonous with
respect to the spectral parameter,
 $$ 
 B_{\kappa}(s,s^{\prime })\leq B_{\kappa ^{\prime }}
 (s,s^{\prime}) \quad \mathrm{for}\quad \kappa ^{\prime }<
 \kappa \,,  
 $$ 
thus to prove lemma it suffices to show that $B_{\kappa _{0}}$ is
a Hilbert-Schmidt operator. Since the function $\upsilon\mapsto
\frac{\e^{-\kappa _{0}\upsilon }}{\upsilon }$ is strictly
decreasing and convex in $(0,\infty)$, we have the following
estimate,
 $$ 
 0\leq \frac{\e^{-\kappa _{0} \rho }}{\rho }-
 \frac{\e^{-\kappa_{0}\sigma }}{\sigma } \leq
 -\left[ \frac{\e^{-\kappa _{0} \sigma_{c}}}{\sigma _{c}}
 \right] ^{\prime}(\sigma -\rho )\,,
 $$ 
where $\sigma_c:= c\sigma$ and $c$ is the constant appearing in
(a1). Thus we get
 \begin{equation}
 0\leq \frac{\e^{-\kappa _{0} \rho }}{\rho }
 -\frac{\e^{-\kappa _{0}\sigma }}{\sigma } \leq
 (\kappa _{0} \sigma _{c}+1)\frac{\sigma
 -\rho }{\sigma _{c}^{2}}\, \e^{-\kappa _{0} \sigma _{c}}\,,
 \label{estim1}
 \end{equation}
and moreover, the assumption (a1) gives the bound
 $$ 
 \frac{\sigma -\rho }{\sigma }\leq 1-c\,.  
 $$ 
In view of (a2), there exists a positive $\tilde{c}$ such that
 $$ 
 \sigma (s,s^{\prime })\geq \tilde{c}\,.  \label{estim3}
 $$ 
holds for any $(s,s^{\prime })\in \R^{2}\setminus S_{\omega
,\varepsilon }$. Combining the last three inequalities we have in
$ \R^{2}\setminus S_{\omega ,\varepsilon }$ the estimate
 \begin{equation}
 \frac{1}{4\pi }\left[\frac{\e^{-\kappa _{0} \rho }}{\rho }-
 \frac{\e^{-\kappa _{0} \sigma }}{\sigma }\right]\leq M_{1}
 \e^{-\kappa _{0} \sigma _{c}} \label{estim4}
 \end{equation}
with $M_{1}:=(4\pi)^{-1}(1\!-\!c)\,c^{-2}(\kappa _{0} c\!+\!
\tilde{c}^{-1})$. On the other hand using (\ref{estim1}) and (a2)
we get
 \begin{equation}
 \frac{1}{4\pi }\left[\frac{\e^{-\kappa _{0} \rho }}{\rho }-
 \frac{\e^{-\kappa _{0}\sigma }}{\sigma }\right]\leq M_{2}
 \e^{-\kappa _{0} \sigma_{c}}\frac{1}{(1+(s^{2}
 \!+\!s^{\prime 2})^{\mu})^{1/2}}  \label{estim5}
 \end{equation}
for $(s,s^{\prime })\in S_{\omega ,\varepsilon }$, where
$M_{2}:=(4\pi)^{-1} dc^{-2} \max \{1,\kappa_{0}c \}$. Putting now
the estimates (\ref{estim4}), (\ref{estim5}) together we find
 \begin{eqnarray*}
 \lefteqn{ \int_{\R^{2}} B_{k}(s,s^{\prime })^{2}dsds^{\prime }}
 \\ && \leq M_{1}^{2}\int_{\R^{2}\setminus S_{\omega,
 \varepsilon}} \e^{-2\kappa _{0}c\left| s-s^{\prime }\right| }
 \,ds\,ds^{\prime} +M_{2}^{2}\int_{S_{\omega ,\varepsilon }}
 \frac{\e^{-2\kappa_{0}c\left| s-s^{\prime }\right| }}
 {1+(s^{2}\!+\!s^{\prime 2})^{\mu}}\,ds\,ds^{\prime } \\ &&\leq
 \frac{1}{2}M_{1}^{2}\, \frac{1+\omega }{1-\omega }\:
 \int_{0}^{\infty } \e^{-\kappa _{0}cu}\,u\,du
 +M_{2}^{2}\int_{S_{\omega,\varepsilon }}
 \frac{\e^{-2\kappa _{0}c\left| s-s^{\prime }\right|}}
 { 1+(s^{2}\!+\!s^{\prime 2})^{\mu }}\,ds\,ds^{\prime }  <\infty\,,
\end{eqnarray*}
which proves the result because the last integral converges for
$\mu>1/2$.
\end{proof} 

\begin{remark} \label{bondBk}
{\rm As we have said, the assumption (a2) includes a decay of the
quantity characterizing the non-straightness at large distances
within $S_{\omega,\varepsilon }$ as well as a restriction for $s$
close to $s^{\prime }$. The latter (which is independent of $\mu
$) ensures the boundedness of $B_{\kappa }$ uniformly w.r.t.
$\kappa$. As in the proof of the above lemma the uniformity is
easy; it suffices to check that $B_{\kappa _{0}}$ is bounded. To
this end we employ the Schur-Holmgren bound: we have $\left\|
B_{\kappa _{0}}\right\| _{l}\leq \left\| B_{\kappa _{0}}\right\|
_{SH}$, where the right-hand side of the last inequality is for
integral operators with symmetric positive kernels defined as
 $$ 
 \left\| B_{\kappa _{0}}\right\| _{SH}=\sup_{s\in \R}\int_{\R}B_{\kappa
 _{0}}(s,s^{\prime })\,ds^{\prime }\,.
 $$ 
Let us use the notation from the previous proof. If $\sigma \leq
\varepsilon $, then by assumption (a2) there exists for any $\mu
\geq 0$ a $C_{1}>0$ such that
$$ 
 B_{\kappa _{0}}(s,s^{\prime })\leq C_{1}.
$$ 
On the other hand, if $\sigma >\varepsilon $ then by
(\ref{estim1}) we can find $C_{2}>0$ such that
$$ 
B_{\kappa _{0}}(s,s^{\prime })\leq C_{2}\,\e^{-\kappa _{0}\sigma
_{c}}.
$$ 
Combining the above two inequalities we get the following
estimate,
$$ 
\int_{\R}B_{\kappa _{0}}(s,s^{\prime })ds^{\prime }\leq
C_{1}\int_{s-\varepsilon }^{s+\varepsilon }ds^{\prime
}+2C_{2}\int_{s+\varepsilon }^{\infty }\e^{-\kappa _{0}c\left|
s-s^{\prime}\right| }ds^{\prime }=2\left( C_{1}\varepsilon
+C_2\frac{\e^{-\kappa _{0}c\varepsilon }}{\kappa _{0}c} \right)
\!,
$$ 
which shows that $\left\| B_{\kappa _{0}}\right\| _{SH}$ is
finite. }
\end{remark}

\begin{lemma} \label{Qkcon9}
Let $\Gamma $ be defined as before. Then the function $\kappa \to
\mathbf{Q} ^{\kappa }$ is continuous in the norm operator in
$(\kappa_0, \infty)$, and moreover,
 \begin{equation}
 \lim_{\kappa \to \infty } \sup\sigma(\mathbf{Q}^{\kappa })
 = -\infty\,.
 \label{-infin}
 \end{equation}
\end{lemma}
\begin{proof}
First we observe that the function $\kappa \to T_{\kappa }$ is
continuous in the norm operator. Indeed, for any $f\in D$ we have
 \begin{eqnarray}
 \left\| (T_{\kappa }\!-\!T_{\kappa ^{\prime }})f\right\| _{l}
 &\!=\!& \frac{1}{4(2\pi)^{3}}
 \int_{\R} \left( \ln \frac{p^{2}+\kappa ^{2}}{p^{2}
 +\kappa ^{\prime 2}}\right) ^{2}|\hat{f}(p)| ^{2}\,dp
 \nonumber \\
 &\!\leq\! &\frac{1}{4(2\pi)^{3}}\left( \ln \frac{\kappa }
 {\kappa^{\prime }}\right) ^{2}\left\| f\right\| _{l}^{2}\to 0.
 \label{conve1}
\end{eqnarray}
as $\kappa ^{\prime }\to \kappa$. On the other hand, in analogy
with \cite{EI} we can estimate
 $$ 
 \left| (B_{\kappa }\!-\!B_{\kappa ^{\prime }})
 (s,s^{\prime})\right| ^{2} \leq 2(B_{\kappa }(s,s^{\prime })^{2}
 \!+\!B_{\kappa^{\prime }}(s,s^{\prime })^{2}) \leq
 4B_{\tilde{\kappa}}(s,s^{\prime })^{2}\,,
 $$ 
where $\tilde{\kappa}:=\min \{\kappa ,\kappa ^{\prime }\}$
arriving therefore at
 \begin{equation}
 \lim_{\kappa ^{\prime }\to \kappa } \left\| B_{\kappa }\!-\!
 B_{\kappa ^{\prime }}\right\| _{HS} \to 0\,;  \label{conve2}
 \end{equation}
from (\ref{conve1}) and (\ref{conve2}) we get the norm-operator
continuity. Let further $f\in D$. The limiting relation
(\ref{-infin}) follows directly from the bound
 \begin{eqnarray*}
 \lefteqn{ (\mathbf{Q}^{\kappa }f,f)_{l} =\frac{1}{(2 \pi )^{3/2}}
 \int_{\R}\left( -\ln \left( p^{2}\!+\!\kappa ^{2}\right) ^{1/2}
 \!+\ln 2 + \psi (1) \right) |\hat{f}(p)|^{2}\, dp }
 \\ &&
 +(B_{\kappa }f,f)_{l} \leq \frac{1}{(2 \pi )^{3/2}}\,
 (-\ln \frac{\kappa }{2} + \psi (1)) \left\| f\right\| _{l}^{2}+
 S\left\| f\right\| _{l}^{2},
 \end{eqnarray*}
where $S:=\sup_{\kappa \geq \kappa _{0}}\|B_{k}\|_{l}<\infty$.
\end{proof} \vspace{1em}

Now we are in position to state and prove our main result.
\begin{theorem} \label{ispo10}
Let $\Gamma $ be a curve determined by a function $\gamma:\:\R\to
\R^3$ which is $C^1$ and piecewise $C^{2}$, and satisfies the
conditions (a1), (a2) with $\mu >1/2$. Then the operator $-\Delta
_{\gamma ,\alpha }$ has at least one isolated eigenvalue in
$(-\infty ,\zeta _{0})$.
\end{theorem}
\begin{proof}
By Lemma~\ref{supsp6} we have $\sup \sigma (\mathbf{Q} ^{\kappa
})>s_{\kappa }$, while by Lemma~\ref{HSnor7} this operator has
only isolated eigenvalues of a finite multiplicity in $(s_{\kappa
},\infty )$. Let $\lambda (\kappa )$ be such an eigenvalue of
$\mathbf{Q}^{\kappa }$. Using then Lemma~\ref{Qkcon9} we conclude
that the function $\lambda (\cdot)$ is continuous and $ \lambda
(\kappa )\to -\infty $ as $\kappa \to \infty $. Consequently,
there is $\tilde{\kappa}>\left| \zeta _{0}\right| ^{1/2}$ such
that $\lambda (\tilde{\kappa})=\alpha $. From the resolvent
formula (\ref{resolv}) we then infer that $-\tilde{\kappa }^{2}\in
(-\infty ,\zeta _{0})$ is an eigenvalue of $-\Delta _{\gamma
,\alpha }$.
\end{proof} \vspace{1em}

\begin{remarks} \label{a2 meaning}
{\rm (a) It is clear that the claim holds without the $C^2$
assumption, however, the latter is needed if we want to interpret
the $\delta$ interaction on the curve in the spirit of
Theorem~\ref{bocon4}. Furthermore, we see that any deviation from
a straight $\Gamma$ pushes the spectrum threshold below the value
$\zeta_0$ but without the assumption (a2) we cannot be sure about
the nature of this added part of the spectrum.
\\ [.5em]
 (b) One may ask what the requirement of
asymptotic straightness expressed by (a2) means. Suppose that
$\gamma $ is $C^{2}$ smooth. Then the curvature of $\Gamma$ is
everywhere defined and can expressed as $k(s)=\left(
\sum_{i=1}^{3}k_{i}(s)^{2}\right) ^{1/2}$, where
$k_{i}(s):=\varepsilon _{ijk}\gamma _{j}^{\prime }(s)\gamma
_{k}^{\prime \prime }(s)$ with the summation convention for the
indices of the Levi-Civita tensor. It allows us estimate the
distance between $\gamma (s)$ and $\gamma (s^{\prime })$ in the
following way,
 \begin{eqnarray*}
 \left| \gamma (s)\!-\!\gamma (s^{\prime })\right| &\!=\!&
 \left[ \sum_{\nu=0}^1
 \left( \sum\nolimits_{i=1}^{3}\int_{s^{\prime }}^{s}
 \cos\left(\int_{s^{\prime }}^{s_{1}}k_{i}(s_{2})\,ds_{2}
 + \frac{\pi\nu}{2} \right)\,ds_{1} \right)^{2} \right]^{1/2}
 \\ &\!\geq\! &\sum\nolimits_{i=1}^{3}\int_{s^{\prime}}^{s}
 \left( 1-\frac{1}{2} \left( \int_{s^{\prime}}^{s_{1}}
 k_{i}(s_{2})\,ds_{2}\right) ^{2}\right)\,ds_{1}\,,
 \end{eqnarray*}
where we assume without loss of generality that $s>s^{\prime }$.
Suppose that there are positive $\beta ,c_{i} $ such that $\left|
k_{i}(s)\right| \leq c_{i}\left| s\right| ^{-\beta }$. Then
$\left| k(s)\right| \leq 3c\left| s\right| ^{-\beta }$, where
$c=\max_{i}\{c_{i}\}$ and one can estimate
 \begin{eqnarray*}
 \lefteqn{ 1-\frac{\left| \gamma (s)\!-\!\gamma(s^{\prime })\right| }
 {\left| s\!-\!s^{\prime }\right| } \le \frac{1}{2\left|
 s\!-\!s^{\prime }\right| }\int_{s^{\prime }}^{s}\left[
 \int_{s^{\prime}}^{s_{1}}k(s_{2})\,ds_{2}\right] ^{2}ds_{1} }
 \\ && \leq \frac{3c^{2}}{2\left| s\!-\!s^{\prime }\right|}\,
 \frac{1}{\left| s^{\prime }\right| ^{2\beta }}
 \int_{s^{\prime}}^{s}\left| s^{\prime }\!-\!s_{1}\right| ^{2}\,ds_{1}
 \le \frac{c^{2}}{2}\,\frac{\left| s^{\prime }\!-\!s\right| ^{2}}{\left|
 s^{\prime }\right| ^{2\beta }} \leq
 \frac{c^{2}}{2}\,\frac{1}{\left| s^{\prime }\right| ^{2\beta -2}}\,.
 \end{eqnarray*}
Thus the conclusion is the same as in the two-dimensional case
discussed in \cite{EI}: the assumption (a2) with $\mu>1/2$ is
satisfied if $\beta>5/4$. }
\end{remarks}


\setcounter{section}{1} \setcounter{equation}{0}
\renewcommand{\theequation}{\Alph{section}.\arabic{equation}}
\renewcommand{\theclaim}{\Alph{section}.\arabic{equation}}
\section*{Appendix: proof of Theorem~\ref{bocon4}}

First we check the inclusion $D(-\Delta _{\gamma , \alpha})
\subseteq \Upsilon_{\alpha }$. Suppose that $f\in D(-\Delta
_{\gamma, \alpha })$, i.e. that there is $ f_{\kappa }\in
D(\Delta)$ such that
 \begin{equation}
 f=f_{\kappa }-\mathbf{\breve{R}}_{\gamma }^{\kappa }(\mathbf{Q}^{\kappa }
 \!-\!\alpha )^{-1}\tau f_{\kappa }\,. \label{fresol}
 \end{equation}
Denote $h:=(\mathbf{Q}^{\kappa }\!-\!\alpha)^{-1}\tau f_{\kappa }
\in L^{2}(\R)$, so $f=f_{\kappa }-\mathbf{R}^{\kappa
}\tau^{\ast}h$. Since $f_{\kappa }\in H^{2}(\R^{3})$ and $\tau
^{\ast }h\in H^{-2}(\R^{3})$ is a measure supported by $\Gamma $
we can conclude that $f\in H_{loc}^{2}(\R^{3}\setminus \Gamma)$ --
see \cite{RS}. Using properties of the Macdonald function
$K_{0}(\varsigma )$ and the following relation
 $$ 
\frac{1}{4\pi } \frac{
 \e^{-\kappa (r^{2}+(s-s^{\prime })^{2})^{1/2}}}
 {(r ^{2}\!+\!(s\!-\!s^{\prime})^{2})^{1/2}}
=\frac{1}{(2\pi )^2}\int _{\R} K_{0}((p_{1}^{2}+\kappa
^{2})^{1/2}r )\, \e^{ip_{1}(s\!-\!s^{\prime})}dp_{1}
$$ 
we can check that
 $$ 
 \lim_{r \to 0}\: \left[\frac{1}{4\pi }\, \int_{\R}\frac{
 \e^{-\kappa (r^{2}+(s-s^{\prime })^{2})^{1/2}}}
 {(r ^{2}\!+\!(s\!-\!s^{\prime})^{2})^{1/2}}
 h(s^{\prime })ds^{\prime }+ \frac {1}{2 \pi } \ln r
 h(s)\right]=T_{\kappa }h(s).
 $$ 
Now it is easy to demonstarte that the function
 $$ 
 (\mathbf{\breve{R}}_{\gamma }^{\kappa }h)(x)=\frac{1}{4\pi}
 \int_{\R}\frac{ \e^{-\kappa \left| x-\gamma (s)\right| }}
 {\left|x\!-\!\gamma (s)\right| }h(s)\,ds
 $$ 
satisfies the limiting relation
 \begin{equation}
 \lim_{r \to 0}\left[ (\mathbf{\breve{R}}^{\kappa }h)\upharpoonright
 _{\Gamma_{r}}(s)+\frac {1}{2 \pi }\ln r h(s) \right] =T_{\kappa }h(s)
 +B_{\kappa}h(s) \label{asympt}
 \end{equation}
with respect to families of ``shifted'' curves described in
Sec.~\ref{bc}.
The above limits are understood in distributional sense. It
follows from (\ref{fresol}) and (\ref{asympt}) that
 \begin{equation}
 \Xi (f)(s)=- \frac {1}{2 \pi } h(s).  \label{logary}
 \end{equation}
On the other hand, since $f_{\kappa }\in D(\Delta )=H^{2}(\R^{3})$
the same relations (\ref{fresol}) and (\ref{asympt}) yield
 \begin{equation}
 \Omega (f)(s)=(\tau f_{\kappa })(s)-(\mathbf{Q}_{\kappa }h)(s)
 =- \alpha h(s). \label{regula}
 \end{equation}
Combining (\ref{logary}) and (\ref{regula}) we obtain that $f\in
\Upsilon $ and $2 \pi \alpha \Xi (f)(s)=\Omega (f)(s)$.
Conversely, one can show by analogous considerations that any
function from $\Upsilon _{\alpha}$ can be represented in the form
$f=f_{\kappa }-\mathbf{\breve{R}}_{\gamma }^{\kappa
}(\mathbf{Q}^{\kappa }-\alpha )^{-1}\tau f_{\kappa }$ with
$f_{\kappa }\in D(\Delta )$, so $D(-\Delta _{\gamma , \alpha
})=\Upsilon _{\alpha }$. Moreover, since
 $$ 
 (-\Delta _{\gamma ,\alpha }+\kappa ^{2})f=(-\Delta +\kappa
 ^{2})f_{\kappa }
 $$ 
and $\tau ^{\ast }h\in H^{-2}(\R^{3})$ is a measure supported
by $\Gamma $ we infer that
 $$ 
-\Delta _{\gamma ,\alpha }f(x)=-\Delta f(x),\quad x\in
\R^{3}\backslash \Gamma\, .
 $$ 
This completes the proof.


\subsection*{Acknowledgments}

The work was supported by GAAS under the contract \#1048101.

\vspace{10mm}

\begin{flushleft}
Department of Theoretical Physics \\ Nuclear Physics Institute \\
Academy of Sciences \\ 25068 \v Re\v z near Prague, Czech Republic
\\ E-mail: exner@ujf.cas.cz, kondej@ujf.cas.cz

\end{flushleft}

\end{document}